\documentclass[aps,preprint,nofootinbib,floatfix,superscriptaddress,showpacs,showkeys]{revtex4}
\usepackage{epsfig}
\usepackage{graphicx}
\usepackage{amsmath}
\usepackage{multirow}
\usepackage{tablefootnote}

\begin{document} 

\title{$\Xi^- t$ quasibound state instead of $\Lambda\Lambda nn$
bound state}
\author{H.~Garcilazo} 
\email{humberto@esfm.ipn.mx} 
\affiliation{Escuela Superior de F\' \i sica y Matem\'aticas,
Instituto Polit\'ecnico Nacional, Edificio 9, 
07738 M\'exico D.F., Mexico} 
\author{A.~Valcarce} 
\email{valcarce@usal.es} 
\affiliation{Departamento de F\'\i sica Fundamental,
Universidad de Salamanca, 37008 Salamanca, Spain}
\author{J. Vijande}
\email{javier.vijande@uv.es}
\affiliation{Departamento de F{\'\i}sica At\'omica, Molecular y Nuclear, 
Universidad de Valencia (UV) and IFIC (UV-CSIC),
46100 Valencia, Spain}
\affiliation{IRIMED Joint Research Unit (IIS La Fe - UV), 46100 Valencia, Spain}

\date{\today} 

\begin{abstract} 
We study the coupled $\Lambda\Lambda nn-\Xi^- pnn$ system 
to check whether the inclusion of channel coupling is able
to bind the $\Lambda\Lambda nn$ system.   
We use a separable potential three-body model 
of the coupled $\Lambda\Lambda nn - \Xi^- pnn$ system as well 
as a variational four-body calculation with realistic interactions.
Our results exclude the possibility of 
a $\Lambda\Lambda nn$ bound state by a large margin. However, we have
found a $\Xi^- t$ quasibound state
above the $\Lambda\Lambda nn$ threshold.
\end{abstract}



\maketitle 

\section{Introduction}
\label{secI}

Bound states of two neutrons and two $\Lambda$ hyperons are a controversial subject.
Recently, Bleser {\it et al.}~\cite{Ble19} have offered a new interpretation 
of the results of the BNL AGS-E906 experiment to produce and study double hypernuclei through
a $(K^-,K^+)$ reaction on $^9$Be~\cite{BNLAGS}.
Following a suggestion made by Avraham Gal, they explored the conjecture
that decays of a $^4_{\Lambda\Lambda}$n double hypernucleus may be 
responsible for some of the
observed structures in the correlated $\pi^- - \pi^-$ momenta.
However, in a recent calculation using the 
stochastic variational method in a pionless effective field theory 
approach~\cite{Cont19}, it has been concluded that the $\Lambda\Lambda nn$
system is unbound by a large margin. We had previously come to the identical 
conclusion~\cite{Gar17} in a study of the uncoupled $\Lambda\Lambda nn$ system
using local central Yukawa-type Malfliet-Tjon interactions reproducing 
the low-energy parameters and phase shifts of the $nn$ system and the 
latest updates of the $n\Lambda$ and $\Lambda\Lambda$ Nijmegen ESC08c potentials.
It is important to notice that in order to create a $\Lambda\Lambda nn$ bound state 
the four particles must coincide simultaneously since the system does not
contain two- or three-body subsystem bound states, so that the probability 
of the event occurring is rather small.

In this work we take the calculation one step further by considering 
the coupled $\Lambda\Lambda nn-\Xi^-pnn$
system to check if the inclusion of channel coupling is able
to bind the $\Lambda\Lambda nn$ system. If this were not the case, we will
study whether there could be a $\Xi^- p nn$ sharp resonance 
or quasibound state above the $\Lambda\Lambda nn$ threshold.  
In the $\Lambda\Lambda nn-\Xi^-pnn$ system, the effect of channel 
coupling arises from the process $\Lambda\Lambda\to \Xi N$ in the 
two-body channel $(i,j)=(0,0)$. The
channel $\Xi N$ can be realized in two ways, $\Xi^0 n$ or $\Xi^- p$;
however, if one restricts the calculation to $S$ waves, the subchannel
$\Xi^0 n$ can not contribute since one can not have three nucleons 
with a symmetric space wave function. Thus, 
only the subchannel $\Xi^- p$ will contribute.\footnote{We have
explicitly checked that the $\Xi^0 nnn$ system is unbound by a large margin due to the mixed
symmetry nature of the spin wave function of the three neutrons, what requires a
mixed symmetry radial wave function. We have calculated the binding energy
of $\Lambda nnn$ and $\Xi^0 nnn$ states with our variational method, obtaining
a result that it is always above threshold. On the other hand, this is reasonable
because if the mixed symmetric radial wave function of the three neutrons
would not penalize the interaction, the $^4$H $\equiv pnnn$ would be bound 
in nature due to the stronger $pn$ interaction. However, no $^4$H positive
parity level has ever been reported.}

We present two different approaches. First, we address
a three-body model $\Lambda\Lambda (nn) - \Xi^- p (nn)$, 
where the dineutron $(nn)$ is treated as a particle of isospin 1 and spin 0, 
and all the two-body interactions are assumed to be simple Yamaguchi 
separable potentials. This allows us to search for solutions
both in the real axis, bound states, and in the complex plane,
resonances and quasibound states. Later, we perform a
variational four-body calculation with realistic local two-body interactions
which necessarily will be restricted to energies in the real axis.  

\section{The $\Lambda\Lambda (nn) - \Xi^- p (nn)$ three-body model}
\label{secII}
In this model we treat the dineutron $(nn)$ as an elementary 
particle with mass $m_{(nn)}=2m_n$, isospin 1, and spin 0 with
two-body interactions given by Yamaguchi separable potentials~\cite{Yam54}. It is based
on the model proposed in Ref.~\cite{Gar16} to search for resonances
of the $\Lambda\Lambda N - \Xi NN$ system. If one of the
nucleons in the lower and upper channels is replaced by a dineutron, $N\to (nn)$, 
the equations of Ref.~\cite{Gar16} are similar to those of this work. The differences
originate from the fact that in the $\Lambda\Lambda N - \Xi NN$ system 
two of the three particles in the upper channel are identical while in the
$\Lambda\Lambda (nn) - \Xi^- p (nn)$ system 
the three particles in the upper channel are different.

\subsection{Three-body equations}
\label{secIIa}
We take the dineutron $(nn)$ as particle 1. 
In the lower channel the two $\Lambda$'s are particles 2 and 3 
while in the upper channel particles 2 and 3 are the $\Xi^-$ 
and $p$, respectively. Following the graphical method of 
Ref.~\cite{Gac14} the equations of the
$\Lambda\Lambda (nn) - \Xi^- p (nn)$ system are,
\begin{eqnarray}
\langle 1|T_1\rangle = &&
2\langle 1|t_1^{\Lambda\Lambda}|1\rangle 
\langle 1|3\rangle G_0(3)\langle 3|T_3\rangle+
\langle 1|t_{1}^{\Lambda\Lambda-\Xi^- p}|1\rangle 
\langle 1|2\rangle G_0(2) \langle 2|U_2\rangle
\nonumber \\ &&
+\langle 1|t_{1}^{\Lambda\Lambda-\Xi^- p}|1\rangle 
\langle 1|3\rangle G_0(3) \langle 3|U_3\rangle,
\nonumber \\
\langle 3|T_3\rangle = &&-\langle 3|t_3^{(nn)\Lambda}|3\rangle 
\langle 2|3\rangle G_0(3)
\langle 3|T_3\rangle+\langle 3|t_3^{(nn)\Lambda}|3\rangle
\langle 3|1\rangle G_0(1)
\langle 1|T_1\rangle, \nonumber \\
\langle 1|U_1\rangle = && \langle 1|t_1^{\Xi^- p}|1\rangle
\langle 1|2\rangle G_0(2)
\langle 2|U_2\rangle +
\langle 1|t_1^{\Xi^- p}|1\rangle
\langle 1|3\rangle G_0(3)
\langle 3|U_3\rangle
\nonumber \\ &&
+2\langle 1|t_1^{\Xi^- p -\Lambda\Lambda}|1\rangle
\langle 1|3\rangle G_0(3)
\langle 3|T_3\rangle,
\nonumber \\
\langle 2|U_2\rangle = && \langle 2|t_2^{(nn)p}|2\rangle 
\langle 2|3\rangle G_0(3)
\langle 3|U_3\rangle+\langle 2|t_2^{(nn)p}|2\rangle
\langle 2|1\rangle G_0(1)
\langle 1|U_1\rangle,
\nonumber \\ 
\langle 3|U_3\rangle = && \langle 3|t_3^{(nn)\Xi^-}|3\rangle 
\langle 3|2\rangle G_0(2)
\langle 2|U_2\rangle+\langle 3|t_3^{(nn)\Xi^-}|3\rangle
\langle 3|1\rangle G_0(1)
\langle 1|U_1\rangle \, .
\label{eq1} 
\end{eqnarray}
For all the uncoupled interactions we assume separable potentials
of the form,
\begin{equation}
V_i^\rho = g_i^\rho\rangle \lambda_i^\rho\langle g_i^\rho \, ,
\label{eq2}
\end{equation}
such that the two-body $t-$matrices are,
\begin{equation}
t_i^\rho = g_i^\rho\rangle \tau_i^\rho\langle g_i^\rho \, ,
\label{eq3}
\end{equation}
with
\begin{equation}
\tau_i^\rho = \frac{1}{({\lambda_i^\rho})^{-1}
-\langle g_i^\rho|G_0(i)|g_i^\rho\rangle} \, .
\label{eq4}
\end{equation}
In the case of the two-body channel responsible for the channel 
coupling, $(i,j)=(0,0)$, we use a separable interaction of the form,
\begin{equation}
V_{1}^{\rho\sigma} = g_1^\rho\rangle \lambda_1^{\rho-\sigma}\langle g_1^\sigma \, ,
\label{eq5}
\end{equation}
such that
\begin{equation}
t_1^{\rho-\sigma} = g_1^\rho\rangle \tau_1^{\rho-\sigma}\langle g_1^\sigma \, ,
\label{eq6}
\end{equation}
with
\begin{eqnarray}
\tau_{1}^{\Lambda\Lambda} &=& 
\frac{-({\lambda_1^{\Lambda\Lambda-\Xi^- p}})^2G^{\Xi^- p}
-\lambda_1^{\Lambda\Lambda}(1-\lambda_1^{\Xi^- p}G^{\Xi^- p})}
{(\lambda_1^{\Lambda\Lambda-\Xi^- p})^2G^{\Lambda\Lambda}G^{\Xi^- p}
-(1-\lambda_1^{\Lambda\Lambda}G^{\Lambda\Lambda})
(1-\lambda_1^{\Xi^- p}G^{\Xi^- p})} \, , \nonumber \\
\tau_{1}^{p\Xi^-} &=& 
\frac{-({\lambda_1^{\Lambda\Lambda-\Xi^- p}})^2G^{\Lambda\Lambda}
-\lambda_1^{\Xi^- p}(1-\lambda_1^{\Lambda\Lambda}G^{\Lambda\Lambda})}
{(\lambda_1^{\Lambda\Lambda-\Xi^- p})^2G^{\Lambda\Lambda}G^{\Xi^- p}
-(1-\lambda_1^{\Lambda\Lambda}G^{\Lambda\Lambda})
(1-\lambda_1^{\Xi^- p}G^{\Xi^- p})} \, , \nonumber \\
\tau_{1}^{\Lambda\Lambda-\Xi^-p} &=& \tau_1^{\Xi^- p -\Lambda\Lambda}=
\frac{-{\lambda_1^{\Lambda\Lambda-\Xi^- p}}}
{(\lambda_1^{\Lambda\Lambda-\Xi^- p})^2G^{\Lambda\Lambda}G^{\Xi^- p}
-(1-\lambda_1^{\Lambda\Lambda}G^{\Lambda\Lambda})
(1-\lambda_1^{\Xi^- p}G^{\Xi^- p})} \, ,
\label{eq69}
\end{eqnarray}
and
\begin{eqnarray}
G^{\Lambda\Lambda} &=& 
\langle g_1^{\Lambda\Lambda}|G_0|g_1^{\Lambda\Lambda}\rangle \, , \nonumber \\
G^{\Xi^- p} &=& 
\langle g_1^{\Xi^- p}|G_0|g_1^{\Xi^- p}\rangle \, ,
\label{eq78}
\end{eqnarray}
where for simplicity we have redefined 
$\tau_1^{\Lambda\Lambda}\equiv
\tau_1^{\Lambda\Lambda-\Lambda\Lambda}$, 
$\lambda_1^{\Lambda\Lambda}\equiv
\lambda_1^{\Lambda\Lambda-\Lambda\Lambda}$, etc.

Using Eqs.~\eqref{eq3} and~\eqref{eq6} into the integral equations~\eqref{eq1} and 
introducing the transformations $\langle i|T_i\rangle =
\langle i|g_i^{\alpha_i}\rangle \langle i|X_i\rangle$ and
$\langle i|U_i\rangle =
\langle i|g_i^{\beta_i}\rangle \langle i|Y_i\rangle$, one
obtains the one-dimensional integral equations
\begin{eqnarray}
\langle 1|X_1\rangle = 
&& 2\tau_1^{\Lambda\Lambda}\langle g_1^{\Lambda\Lambda}|1\rangle 
\langle 1|3\rangle G_0(3)\langle 3|g_3^{(nn)\Lambda}\rangle
\langle 3|X_3\rangle
\nonumber \\ &&
+ \tau_1^{\Lambda\Lambda-\Xi^- p}\langle g_1^{\Xi^- p}|1\rangle 
\langle 1|2\rangle G_0(2) \langle 2|g_2^{(nn)p}\rangle
\langle 2|Y_2\rangle
\nonumber \\ &&
+ \tau_1^{\Lambda\Lambda-\Xi^- p}\langle g_1^{\Xi^- p}|1\rangle 
\langle 1|3\rangle G_0(3) \langle 3|g_3^{(nn)\Xi^-}\rangle
\langle 3|Y_3\rangle \, ,
\nonumber \\
\langle 3|X_3\rangle = && -\tau_3^{(nn)\Lambda}
\langle g_3^{(nn)\Lambda}|3\rangle\langle 2|3\rangle G_0(3)
\langle 3|g_3^{(nn)\Lambda}\rangle
\langle 3|X_3\rangle
\nonumber \\ &&
+ \tau_3^{(nn)\Lambda}\langle g_3^{(nn)\Lambda}|3\rangle
\langle 3|1\rangle G_0(1)
\langle 1|g_1^{\Lambda\Lambda}\rangle
\langle 1|X_1\rangle \, ,
\nonumber \\
\langle 1|Y_1\rangle = && \tau_1^{\Xi^- p}\langle g_1^{\Xi^- p}|1\rangle
\langle 1|2\rangle G_0(2)
\langle 2|g_2^{(nn)p}\rangle\langle 2|Y_2\rangle 
\nonumber \\ &&
+\tau_1^{\Xi^- p} \langle g_1^{\Xi^- p}|1\rangle
\langle 1|3\rangle G_0(3)
\langle 3|g_3^{(nn)\Xi^-}\rangle\langle 3|Y_3\rangle \nonumber \\ &&
+2\tau_1^{\Xi^- p -\Lambda\Lambda}\langle g_1^{\Lambda\Lambda}|1\rangle
\langle 1|3\rangle G_0(3)
\langle 3|g_3^{(nn)\Lambda}\rangle\langle 3|X_3\rangle \, ,
\nonumber \\
\langle 2|Y_2\rangle = && \tau_2^{(nn)p}\langle g_2^{(nn)p}|2\rangle 
\langle 2|3\rangle G_0(3)
\langle 3|g_3^{(nn)\Xi^-}\rangle
\langle 3|Y_3\rangle
\nonumber \\ &&
+ \tau_2^{(nn)p}\langle g_2^{(nn)p}|2\rangle
\langle 2|1\rangle G_0(1)
\langle 1|g_1^{\Xi^- p}\rangle
\langle 1|Y_1\rangle \, ,
\nonumber \\ 
\langle 3|Y_3\rangle = && \tau_3^{(nn)\Xi^-}\langle g_3^{(nn)\Xi^-}|3\rangle 
\langle 3|2\rangle G_0(2)
\langle 2|g_2^{(nn)p}\rangle
\langle 2|Y_2\rangle
\nonumber \\ &&
+ \tau_3^{(nn)\Xi^-}\langle g_3^{(nn)\Xi^-}|3\rangle
\langle 3|1\rangle G_0(1)
\langle 1|g_1^{\Xi^- p}\rangle
\langle 1|Y_1\rangle \, .
\label{eq9} 
\end{eqnarray}
Eqs.~\eqref{eq9} can be extended into the complex energy plane 
following the method of Ref.~\cite{Afn74}.

\subsection{Two-body inputs}
\label{secIIb}
The $\Xi^-t\to \Lambda\Lambda nn$ process 
occurs with quantum numbers $(I,J)=(1,0)$ so that,
since we restrict our calculation to $S$ waves,
the contributing two-body channels in our three-body model are: the
$(nn)p$ channel $(i,j)=(1/2,1/2)$, the 
$(nn)\Lambda$ channel  $(i,j)=(1,1/2)$, the 
$(nn)\Xi^-$ channel $(i,j)=(3/2,1/2)$, and the $\Lambda\Lambda - \Xi^- p$
channel $(i,j)=(0,0)$.

We use Yamaguchi form factors for the separable potentials of
Eqs.~\eqref{eq2} and~\eqref{eq5}, i.e.,
\begin{equation}
g(p)=\frac{1}{\alpha^2+p^2} \, .
\label{eq11}
\end{equation}
Thus, for each uncoupled two-body channel we have to fit the
two parameters $\alpha$ and $\lambda$.

In the case of the $(nn)p$ subsystem with quantum numbers
$(i,j)=(1/2,1/2)$, the tritium channel, for a given value of the range
$\alpha$ the tritium binding
energy ($8{.}48$ MeV) determines the strength $\lambda$
through Eq.~\eqref{eq4} as,
\begin{equation}
\lambda=\frac{1}{\langle g|G_0(E_B))|g\rangle} \, ,
\label{eq12}
\end{equation}
while the value of $\alpha$ is determined from the binding energy of 
$^4{\rm He}$ (28.2 MeV) through the solution of the three-body system $(nn)pp$.
The parameters of this model are given in Table~\ref{tab1}.
\begin{table}[t]
\caption{Parameters of the different separable potential models
for the uncoupled partial waves: $\alpha$ (in fm$^{-1}$) and $\lambda$
(in fm$^{-2}$).}
\begin{ruledtabular} 
\begin{tabular}{cccccc} 
Model & Subsystem  & $(i,j)$ &  $\alpha$ & $\lambda$  & \\
\hline
& $(nn)p$                                 &  (1/2,1/2)      & 1.07   & $-$0.5444 & \\
\multirow{2}{*}{1} & $(nn)\Lambda$        &  (1,1/2)        & \multirow{2}{*}{1.0}    & $-$0.1655 & \\
                   & $(nn)\Xi^-  $        &  (3/2,1/2)      &                         & $-$0.2904 & \\
\multirow{2}{*}{2} & $(nn)\Lambda$        &  (1,1/2)        & \multirow{2}{*}{2.0}    & $-$1.1560 & \\
                   & $(nn)\Xi^-  $        &  (3/2,1/2)      &                         & $-$1.7719 & \\
\multirow{2}{*}{3} & $(nn)\Lambda$        &  (1,1/2)        & \multirow{2}{*}{3.0}    & $-$3.9450 & \\
                   & $(nn)\Xi^-  $        &  (3/2,1/2)      &                         & $-$5.4162 & 
									\end{tabular}
\end{ruledtabular}
\label{tab1} 
\end{table}

In the case of the $(nn)\Lambda$ subsystem with quantum numbers
$(i,j)=(1,1/2)$, we fit the two parameters of the interaction
to the ground state and spin-excitation energies of the
$^4_\Lambda{\rm H}$ hypernucleus. It is considered as a three-body
system $(nn)p\Lambda$ with quantum numbers $(I,J)=(1/2,0)$. For
the $(nn)p$ subsystem we use the interaction previously described and
for the $p\Lambda$ the separable potentials for $j=0$ and $j=1$
constructed in Ref.~\cite{Gar16}. Thus,
for a given value of the range $\alpha$,
we fit the strength $\lambda$ to the binding energy 
of $^4_\Lambda{\rm H}$ (10.52 MeV)~\cite{Gal16}.
In order to obtain the range $\alpha$ we calculate the 
binding energy of the excited state $(I,J)=(1/2,1)$ (9.43 MeV)~\cite{Gal16} obtaining
for $\alpha$ = 1, 2, and 3 fm$^{-1}$ the values 9.93, 9.81 and 9.77 MeV,
respectively, which are labeled as models 1, 2, and 3 in Table~\ref{tab1}. As it is
well known, the $^4_\Lambda{\rm H}$ spin excitation is difficult to fit since it depends
strongly on the tensor force arising from the transition 
$\Lambda N - \Sigma N$~\cite{Hiy14,Gal14,Gal16,Gar14}. Therefore,
we did not consider larger values of $\alpha$.

In the case of the $(nn)\Xi^-$ subsystem with quantum numbers
$(i,j)=(3/2,1/2)$, we do not have any experimental information 
available to calibrate our separable potential model. However,
in a couple of recent calculations~\cite{Gac16,Fil17} 
based in the strangeness $-$2 Nijmegen ESC08c potential~\cite{Nag15}
a bound state is predicted with a binding energy of 2.89 MeV
below the $\Xi NN$ threshold.
Thus, we have used this result to obtain the strength $\lambda$ of the
separable potential using Eq.~\eqref{eq12} and
taking the range $\alpha$ equal to that of the
$(nn)\Lambda$ subsystem. We give in Table~\ref{tab1} the parameters 
corresponding to the different models 1, 2, and 3.
\begin{table}[t]
\caption{Parameters of the two separable potential models
for the coupled partial wave $(i,j)=(0,0)$: $\alpha_1^{\Lambda\Lambda}$, 
$\alpha_1^{\Xi N}$ (in fm$^{-1}$), $\lambda_1^{\Lambda\Lambda}$,
$\lambda_1^{\Xi N}$, and $\lambda_1^{\Lambda\Lambda - \Xi N}$
(in fm$^{-2}$).} 
\begin{ruledtabular} 
\begin{tabular}{cccccccc} 
& Model  &    $\alpha_1^{\Lambda\Lambda}$ & $\lambda_1^{\Lambda\Lambda}$
& $\alpha_1^{\Xi N}$  & $\lambda_1^{\Xi N}$  & $\lambda_1^{\Lambda\Lambda - \Xi N}$
  &     \\
\hline
& A & 1.3465 & $-$0.1390 &   1.1460 & $-$0.3867 & 0.0977  & \\
& B & 1.25 & $-$0.0959 & 4.287 & 1.302 & 1.243  & \\
\end{tabular}
\end{ruledtabular}
\label{tab2} 
\end{table}
\begin{table}[b]
\caption{Energy eigenvalue of the $\Lambda\Lambda (nn) - \Xi^- p (nn)$
system (in MeV) measured with respect to the $\Xi^-pnn$ threshold.
The results in parenthesis are those of the uncoupled $\Xi^- t$ binding energy.}
\begin{ruledtabular} 
\begin{tabular}{cccccc} 
& Model  & 1 &  2 & 3  & \\
\hline
& A & $-12{.}80 - i\, 0{.}05$ ($-$12.73)  & $-13{.}46 - i\, 0{.}04$ ($-$13.37)    & $-13{.}52 - i\, 0.04$ ($-$13.43) & \\
& B & $-10{.}99 - i\, 0{.}06$ ($-$10.92)  & $-11{.}04 - i\, 0{.}07$ ($-$10.93)    & $-10{.}89 - i\, 0.07$ ($-$10.77) & \\
\end{tabular}
\end{ruledtabular}
\label{tab3} 
\end{table}

In the case of the coupled $\Lambda\Lambda - \Xi^-p$ subsystem first we use
a recent lattice QCD study by the HAL QCD Collaboration~\cite{Sas18} with almost
physical quark masses ($m_\pi=146$ MeV and $m_K=525$ MeV). 
In this model the $H$ dibaryon was calculated through the coupled 
channel $\Lambda\Lambda - \Xi N$ system, appearing
as a very sharp resonance just below the $\Xi N$ threshold~\cite{Sas18,Ino12}. 
We have constructed a model,
labeled as A, giving similar $\Lambda\Lambda$  and $\Xi N$ phase shifts as
those of Ref.~\cite{Sas18}. The 
parameters of this model are given in Table~\ref{tab2}. Besides, we have also
considered the separable potential model of the $\Lambda\Lambda-\Xi N$ system
constructed in Ref.~\cite{Gar16} which is based in the Nijmegen
ESC08c potential~\cite{Nag15}. This model is shown in Table~\ref{tab2} as model B. Of
course, in the $\Lambda\Lambda (nn) - \Xi^- p (nn)$ calculation we use the parameters 
$\lambda_1^{\Lambda\Lambda-\Xi^-p}=\lambda_1^{\Lambda\Lambda-\Xi N}/\sqrt{2}$
and $\lambda_1^{\Xi^-p}=\lambda_1^{\Xi N}/2$.

\subsection{Results}
\label{secIIc}
We show in Table~\ref{tab3} the energy eigenvalue of the two models A$-$B
of the coupled $\Lambda\Lambda - \Xi N$ system 
and the three models 1$-$3 of the $(nn)\Lambda$ and $(nn)\Xi^-$ systems. We also 
give in parentheses the energy of the uncoupled $\Xi^- t$ system. As one
can see from this table, the real part of the energy eigenvalue is
slightly below the energy of the uncoupled $\Xi^- t$ system and 
the imaginary part of the energy eigenvalue is roughly the 
difference between the uncoupled energy and the real part of 
the energy eigenvalue. Thus, this state appears as a narrow $\Xi^- t$
quasibound state decaying to $\Lambda\Lambda nn$. The reason 
for the narrow width of the $\Xi^- t$ state stems from the weakness of the
$\Lambda\Lambda - \Xi N$ transition potential~\cite{Nag15,Sas18}, that on the other hand is also
responsible for the $H$ dibaryon appearing as a 
very sharp resonance just below the $\Xi N$ threshold~\cite{Ino12}.

Finally, we give in Table~\ref{tab3p} the corresponding values of the
$\Xi^-t$ scattering lengths of the two models A$-$B which may be of use
in the calculation of the energy shift of the 
atomic levels of the $\Xi^-t$ atom.
\begin{table}[t]
\caption{$\Xi^-t$ scattering length (in fm).
The results in parenthesis are those of the uncoupled $\Xi^- t$ scattering length.}
\begin{ruledtabular} 
\begin{tabular}{cccccc} 
& Model  & 1 &  2 & 3  & \\
\hline
& A & $1{.}286 - i\, 0{.}005$ $(1{.}293)$  & $1{.}030 - i\, 0{.}003$ $(1{.}036)$    & $0{.}957 - i\, 0{.}003$ $(0{.}963)$ & \\
& B & $1{.}551 - i\, 0{.}015$ $(1{.}567)$  & $1{.}315 - i\, 0{.}016$ $(1{.}339)$    & $1{.}268 - i\, 0{.}018$ $(1{.}298)$ & \\
\end{tabular}
\end{ruledtabular}
\label{tab3p} 
\end{table}

\section{The $\Lambda\Lambda nn$ and $\Xi^-pnn$ four-body problems}
\label{secIII}
\subsection{Four-body calculation}
\label{secIIIa}
The four-body problem has been addressed by means of a Generalized Gaussian Variational (GGV)
method~\cite{Vij09,Via09}. The nonrelativistic Hamiltonian is given by,
\begin{equation}
H=\sum_{i=1}^4\frac{\vec p_{i}^{\,2}}{2m_{i}}+\sum_{i<j=1}^4V_{ij}(\vec r_{ij}) \, ,
\label{ham}
\end{equation}
where $V(\vec r_{ij})$ is a local central two-body potential.

The four-body wave function 
is taken to be a sum over all allowed channels with well-defined symmetry properties: 
\begin{equation}
\psi(\vec x,\vec y,\vec z) = \sum^{s}_{\kappa=1} \chi^{SI}_{\kappa} \, 
R_{\kappa}(\vec x,\vec y,\vec z),
\label{trial-wave-function}
\end{equation}
where $s$ is the number of channels allowed by the Pauli principle. $\vec x = \vec r_1 - \vec r_2$, 
$\vec y = \vec r_3 - \vec r_4$, and $\vec z =(m_1 \vec r_1 + m_2 \vec r_2)/(m_1 + m_2) -
(m_3 \vec r_3 + m_4 \vec r_4)/(m_3 + m_4)$ are the Jacobi coordinates. $\chi^{SI}_\kappa$ are orthonormalized 
spin-isospin vectors and $R_{\kappa}(\vec x,\vec y,\vec z)$ is the radial part 
of the wave function of the $\kappa^\text{th}$ channel. 
In order to get the appropriate symmetry properties
in configuration space, $R_{\kappa}(\vec x,\vec y,\vec z)$ 
is expressed as the sum of four components, 
\begin{equation}
R_{\kappa}(\vec x,\vec y,\vec z) = \sum^4_{n=1} w_\kappa^n \, R^n_\kappa(\vec x,\vec y,\vec z) ,
\label{Rk}
\end{equation}
where $w_\kappa^n = \pm 1$. Finally, each $ R^n_\kappa(\vec x,\vec y,\vec z)$ is expanded in terms 
of $N$ generalized Gaussians
\begin{eqnarray}
R^n_\kappa(\vec x,\vec y,\vec z) &=& \nonumber\\
\sum^{N}_{i=1} \alpha^i_{\kappa}
&&\!\!\!\!\!\!\!\! \exp\left[ - a^{i}_\kappa\,\vec x^{\, 2} - b^{i}_\kappa \,\vec y^{\, 2} - c^{i}_\kappa \,\vec z^{\, 2}
- d^{i}_\kappa \, s_1^n \,\vec x\cdot\vec y - e^{i}_\kappa \, s_2^n \,\vec x\cdot\vec z
- f^{i}_\kappa \, s_3^n \,\vec y\cdot\vec z\right] ,
\label{Rkr}
\end{eqnarray}
where $s_i^n$ are equal to $\pm 1$ to guarantee the symmetry properties of the radial wave
function and $\alpha^{i}_\kappa,a^{i}_\kappa,\cdots, f^{i}_\kappa$ are the variational parameters. 
The latter are determined by minimizing the intrinsic energy of the four-body system. We follow closely
the developments of Refs.~\cite{Vij09,Via09}, where further technical details can be found about the 
wave function and the minimization procedure. 
\begin{table}[b]
\caption{$S$ wave two-body channels contributing to the $nnpp$ system with $(I,J)=(0,0)$.}
\begin{center} 
\begin{tabular}{|cccp{0.3cm}|p{0.3cm}ccc|}
  \hline\hline
	$V_{12}$                      & $-$ & $V_{34}$            & & &
	$V_{13}$                      & $-$ & $V_{24}$\\ \hline
	$nn$ $(i,j)=(1,0)$            & $-$ & $pp$ $(i,j)=(1,0)$ & & &
	$np$ $(i,j)=(0,1)$            & $-$ & $np$ $(i,j)=(0,1)$ \\	
	                              &     &                    &  & &
	$np$ $(i,j)=(1,0)$            & $-$ & $np$ $(i,j)=(1,0)$\\
	\hline\hline
 \end{tabular}
\end{center} 
\label{tabx1}
 \end{table}

The numerical method described in this section has been tested
in different few-body calculations in comparison to the hyperspherical harmonic 
formalism, see for example Refs.~\cite{Via09,Vin09}, or the 
stochastic variational approach of Ref.~\cite{Suz98} for some of
the results presented in Ref.~\cite{Vij16}.
As a benchmark calculation to show the capability of the method we have studied the $^4{\rm He}$,
a $nnpp$ system with $(I,J)=(0,0)$, using the spin-averaged Malfliet-Tjon (MT-V) potential of Ref.~\cite{Mal69}. 
Results for the $(I,J)=(0,0)$ four-nucleon problem can be found in Table 11.2 of Ref.~\cite{Suz98}. 
It was solved with different numerical methods getting a full converged binding energy of $31{.}3$ MeV.

We have studied the $(I,J)=(0,0)$ $nnpp$ state with the GGV method
using the MT-V potential of Ref.~\cite{Mal69},
\begin{equation}
V_{ij}(r)=-A\frac{e^{-\mu_Ar}}{r}+B\frac{e^{-\mu_Br}}{r} \, ,
\label{eq21} 
\end{equation}
with parameters: $A=578.09$ MeV, $\mu_A=1{.}55$ fm$ ^{-1}$,
$B=1458.05$ MeV, $\mu_B=3{.}11$ fm$ ^{-1}$. As in Ref.~\cite{Suz98} we have
used $\hbar^2 / m_N = 41{.}47$ MeV fm$^2$.
Being a pure $S$ wave calculation, the different two-body channels
contributing to the $(I,J)=(0,0)$ $nnpp$ state are shown in Table~\ref{tabx1}.
With $N=25$ generalized Gaussians in Eq.~\eqref{Rkr} we have obtained a binding energy of 
$31{.}2$ MeV, which shows the capability of our method and gives confidence
in the results. Let us note that the spin-averaged MT-V potential 
reproduces reasonably well the tritium binding energy, giving a result of 
$8{.}25$ MeV.
 
\subsection{The $\Lambda\Lambda nn$ system}
\label{secIIIb}
\begin{table}[t]
\caption{$S$ wave two-body channels contributing to the $\Lambda\Lambda nn$ system with $(I,J)=(1,0)$.}
\begin{center} 
\begin{tabular}{|cccp{0.3cm}|p{0.3cm}ccc|}
  \hline\hline
	$V_{12}$                      & $-$ & $V_{34}$                       & & &
	$V_{13}$                      & $-$ & $V_{24}$\\ \hline
	$nn$ $(i,j)=(1,0)$            & $-$ & $\Lambda\Lambda$ $(i,j)=(0,0)$ & & &
	$n\Lambda$ $(i,j)=(1/2,0)$    & $-$ & $n\Lambda$ $(i,j)=(1/2,0)$ \\	
	                              &     &                                &  & &
	$n\Lambda$ $(i,j)=(1/2,1)$    & $-$ & $n\Lambda$ $(i,j)=(1/2,1)$ \\	
	\hline\hline
 \end{tabular}
\end{center} 
\label{tabx2}
 \end{table}
\begin{table}[b]
\caption{Low-energy parameters and parameters of the local central Yukawa-type potentials 
given by Eq.~\eqref{eq21} for the $NN$, $\Lambda N$, and $\Lambda \Lambda$ systems
contributing to the $(I,J)=(1,0)$ $nn\Lambda\Lambda$ state.} 
\begin{ruledtabular} 
\begin{tabular}{cccccccccc} 
& Ref. & $(i,j)$ & $A$(MeV fm) & 
$\mu_A({\rm fm}^{-1}$) 
& $B$(MeV fm) & $\mu_B({\rm fm}^{-1})$  & $a({\rm fm})$ & $r_0({\rm fm})$ & \\
\hline
\multirow{1}{*}{$NN$} & ~\cite{Gib90} & $(1,0)$  &  $513.968$  & $1.55$  & $1438.72$ & $3.11$ & $-23.56$ & $2.88$ &\\
\multirow{2}{*}{$\Lambda N$}& \multirow{2}{*}{~\cite{Nae15}} & $(1/2,0)$ &   $416$  & $1.77$  & $1098$ & $3.33$ & $-2.62$  & $3.17$  &\\ 
&& $(1/2,1)$ &   $339$  & $1.87$  & $968$ & $3.73$ & $-1.72$  & $3.50$  &\\ \hline
\multirow{2}{*}{$\Lambda \Lambda$} & ~\cite{Nag15} & $(0,0)$ &   $121$  & $1.74$  & $926$ & $6.04$ & $-0.85$& $5.13$ &\\ 
 & ~\cite{Sas18} & $(0,0)$ &   $207.44$  & $1.87$  & $627.6$ & $3.63$ & $-0.62$& $7.32$ &\\
\end{tabular}
\end{ruledtabular}
\label{t1} 
\end{table}

The uncoupled $\Lambda\Lambda nn$ system with $(I,J)=(1,0)$ was examined in detail in Ref.~\cite{Gar17}
using local central Yukawa-type Malfliet-Tjon interactions. We summed up in Table~\ref{tabx2}
the different two-body channels contributing to the $(I,J)=(1,0)$ $\Lambda\Lambda nn$ state.
The parameters of the $\Lambda N$ and $\Lambda \Lambda$ two-body 
channels were obtained by fitting the low-energy data and the phase-shifts of each channel as given 
in the most recent update of the strangeness $-1$~\cite{Nae15} and $-2$~\cite{Nag15} Nijmegen ESC08c
potential. The low-energy data and the parameters of these models, together
with those of the $NN$ interaction from Ref.~\cite{Gib90}, are given in Table~\ref{t1}.
As can be seen in Fig. 2 of Ref.~\cite{Gar17} there is no $\Lambda\Lambda nn$ bound state. 

The system hardly gets bound for a reasonable increase of the strength of
the $\Lambda\Lambda$ interaction. 
Although one cannot exclude that the genuine $\Lambda\Lambda$ interaction in 
dilute states as the one studied here could be slightly stronger that the one 
reported in Ref.~\cite{Nag15}, however, one needs a multiplicative factor
in the attractive term of Eq.~\eqref{eq21} $g_{\Lambda\Lambda} \ge 1.8$ to
get a bound state. Such modification would destroy the agreement with the Nijmegen ESC08c 
$\Lambda\Lambda$ phase shifts. Note also that this is a very sensitive 
parameter for the study of double-$\Lambda$ hypernuclei~\cite{Nem03}
and this modification would produce an almost $\Lambda\Lambda$ bound state
in free space, in particular it would give rise to $a_{^1S_0}^{\Lambda\Lambda} = -29.15$ fm
and ${r_0}_{^1S_0}^{\Lambda\Lambda} = 1.90$ fm. 
The four-body system would also become bound by taking a multiplicative factor $1.2$ in the $NN$ interaction.
However, such a change would make the $^1S_0$ $NN$ potential as strong as the $^3S_1$~\cite{Mal69} and 
thus the singlet $S$ wave would develop a dineutron bound state, $a_{^1S_0}^{NN} = 6.07$ fm
and ${r_0}_{^1S_0}^{NN} = 1.96$ fm.
The situation is slightly different when dealing with the $\Lambda N$ interaction. 
We have used a common factor $g_{N\Lambda}$ for the attractive part of the two $\Lambda N$
partial waves, $^1S_0$ and $^3S_1$. 
The four-body system develops a bound state for $g_{N\Lambda}=1.1$,
giving rise to the $\Lambda N$ low-energy parameters: $a_{^1S_0}^{\Lambda N} = -5.60$ fm,
${r_0}_{^1S_0}^{\Lambda N} = 2.88$ fm, $a_{^3S_1}^{\Lambda N} = -2.91$ fm, and
${r_0}_{^3S_1}^{\Lambda N} = 2.99$ fm, far from the values constrained by the
existing experimental data. In particular, these scattering lengths point to the
unbound nature of the $\Lambda\Lambda nn$ system based on the hyperon-nucleon
interactions derived from chiral effective field theory in Ref.~\cite{Hai19}, because it is
less attractive: $a_{^1S_0}^{\Lambda p} \in [-2.90, -2.91]$ fm and  
$a_{^3S_1}^{\Lambda p} \in [-1.40,-1.61]$ fm (see Table 1 of Ref.~\cite{Hai19}).

It is also worth mentioning that Ref.~\cite{Ric15} tackled the same problem by fitting 
low-energy parameters of older versions of the
Nijmegen-RIKEN potential~\cite{Rij10,Rij13} or chiral effective field 
theory~\cite{Pol07,Hai13}, by means of
a single Yukawa attractive term or a Morse parametrization. 
The method used to solve the four-body problem is similar to the one
we have used in our calculation, thus the results might be directly 
comparable. Our improved description of the two- and three-body subsystems
and the introduction of the repulsive barrier for the $^1S_0$ $NN$ partial wave,
relevant for the study of the triton binding energy (see Table II 
of Ref.~\cite{Mal70}), leads to a four-body state above threshold,
that cannot get bound by a reliable modification of the
two-body subsystem interactions. As clearly explained in Ref.~\cite{Ric15}, the window of
Borromean binding is more and more reduced for potentials 
with harder inner cores.

For the sake of consistency with Sec.~\ref{secII} we have
repeated the calculation using the latest $\Lambda\Lambda$ interaction derived by the 
HAL QCD Collaboration~\cite{Sas18}. The parameters of the $\Lambda\Lambda$ HAL QCD potential
are given in the last line of Table~\ref{t1}. Although the $\Lambda\Lambda$ interaction
of Ref.~\cite{Sas18} is slightly more attractive than that of the 
Nijmegen ESC08c potential~\cite{Nag15}, the $\Lambda \Lambda nn$ state remains unbound.
The more attractive character of the HAL QCD $\Lambda\Lambda$ interaction can be easily tested
by trying to generate a $\Lambda\Lambda nn$ bound state with the multiplicative factor
in the attractive term of Eq.~\eqref{eq21} of the $\Lambda\Lambda$ interaction.
While with the model of Ref.~\cite{Nag15} a multiplicative factor 
$g_{\Lambda\Lambda}=1.8$ is necessary to get a bound state, with that of Ref.~\cite{Sas18} the bound 
state is developed for $g_{\Lambda\Lambda}=1.6$.

We have also studied the coupled $\Lambda\Lambda nn - \Xi^- p nn$ system, to check if the coupling
to the upper channel $\Xi^- pnn$ could help to generate a $\Lambda\Lambda nn$ bound state. For this purpose
one needs a parametrization of the $\Lambda\Lambda - \Xi N$ transition potential.
As has been explained in Sec.~\ref{secIIb} the HAL QCD Collaboration has recently derived
a $\Lambda\Lambda - \Xi N$ transition potential~\cite{Sas18} with almost physical quark 
masses. In their results, the $H$ dibaryon appears as a very sharp resonance just
below the $\Xi N$ threshold, what points to a rather weak $\Lambda\Lambda - \Xi N$ transition potential. 
We have parametrized this interaction by means of a Malfliet-Tjon interaction as in Eq.~\eqref{eq21} 
with parameters: $A=61{.}66$ MeV, $\mu_A=1{.}79$ fm$ ^{-1}$,
$B=227{.}01$ MeV, $\mu_B=3{.}25$ fm$ ^{-1}$. The details of the $\Xi N$ interaction 
are discussed in the next subsection. The coupled $\Lambda\Lambda nn - \Xi^- p nn$ 
system is clearly unbound. Thus, it only remains to study the possible existence of 
a $\Xi^- p nn$ bound state that may decay to $\Lambda\Lambda nn$.

\subsection{The $\Xi^- p nn$ system}
\label{secIIIc}
We now study the uncoupled $\Xi^- p nn$ system with quantum numbers
$(I,J)=(1,0)$, to look for a possible bound state. This system contains several 
bound states made of subsets of two- and three-body particles. It contains the deuteron, the tritium,
the $(i,j)=(1,1)$ $\Xi N$ bound state predicted by the Nijmegen potential~\cite{Nag15}
with a binding energy of $1{.}56$ MeV, and the $(i,j)=(3/2,1/2)$ $\Xi NN$ bound state with a
binding energy of $2{.}89$ MeV discussed in Sec.~\ref{secIIb}. If there is a $\Xi^- p nn$ bound state, 
it would not be stable unless its binding energy exceeds $m_{\Xi^- p} - m_{\Lambda\Lambda} = 28{.}6$
MeV. Otherwise it would decay to $\Lambda\Lambda nn$. If its binding energy would be larger
than that of the tritium, it would appear as a $\Xi^- t$ resonance or quasibound state
decaying to $\Lambda\Lambda nn$. 

\begin{table}[t]
\caption{$S$ wave two-body channels contributing to the $\Xi^- p nn$ system with $(I,J)=(1,0)$.}
\begin{center} 
\begin{tabular}{|cccp{0.3cm}|p{0.3cm}ccc|}
  \hline\hline
	$V_{12}$                      & $-$ & $V_{34}$               & & &
	$V_{13}$                      & $-$ & $V_{24}$\\ \hline
	$nn$ $(i,j)=(1,0)$            & $-$ & $p\Xi^-$ $(i,j)=(0,0)$ & & &
	$np$ $(i,j)=(1,0)$            & $-$ & $n\Xi^-$ $(i,j)=(1,0)$ \\	
	$nn$ $(i,j)=(1,0)$            & $-$ & $p\Xi^-$ $(i,j)=(1,0)$ & & &
	$np$ $(i,j)=(0,1)$            & $-$ & $n\Xi^-$ $(i,j)=(1,1)$\\
	\hline\hline
 \end{tabular}
\end{center} 
\label{tabx3}
 \end{table}
\begin{table}[b]
\caption{Low-energy parameters and parameters of the local central Yukawa-type potentials 
given by Eq.~\eqref{eq21} for the $\Xi N$ system
contributing to the $(I,J)=(1,0)$ $\Xi^- p nn$ state.}
\begin{ruledtabular} 
\begin{tabular}{cccccccccc} 
& Ref. & $(i,j)$ & $A$(MeV fm) & $\mu_A({\rm fm}^{-1}$) 
& $B$(MeV fm) & $\mu_B({\rm fm}^{-1})$  & $a({\rm fm})$ & $r_0({\rm fm})$ & \\
\hline
\multirow{4}{*}{$\Xi N$}
&~\cite{Sas18} & \multirow{2}{*}{$(0,0)$}              &  $161.38$  & $1.17$  & $197.5$   & $2.18$ & $-$ & $-$ &\\
&~\cite{Nag15} &                                       &  $120$     & $1.30$  & $510$     & $2.30$ & $-$ & $-$ &\\
&~\cite{Nag15} & $(1,0)$                               &  $290$     & $3.05$  & $155$     & $1.60$ & $0.58$  & $-2.52$ &\\
&~\cite{Nag15} & $(1,1)$                               &  $568$     & $4.56$  & $425$     & $6.73$ & $4.91$  & $0.53$  & 
\end{tabular}
\end{ruledtabular} 
\label{tabx4}
\end{table}

To perform this study we need the $\Xi N$ in three different partial waves.
We show in Table~\ref{tabx3} the different two-body channels 
contributing to the $(I,J)=(1,0)$ $\Xi^- p nn$ state.
Firstly, we use the full set of $\Xi N$ interactions of the Nijmegen group~\cite{Nag15}.
As in the case of the two-body channels in Sec.~\ref{secIIIb},
we have constructed the two-body amplitudes for all subsystems entering the four-body 
problem studied by solving the Lippmann--Schwinger
equation of each $(i,j)$ channel,
\begin{equation}
t_{ij}(p,p';e)= V_{ij}(p,p')+\int_0^\infty {p^{\prime\prime}}^2
dp^{\prime\prime} V_{ij}(p,p^{\prime\prime})
\frac{1}{e-{p^{\prime\prime}}^2/2\mu} t_{ij}(p^{\prime\prime},p';e) \, ,
\label{eq19} 
\end{equation}
where 
\begin{equation}
V_{ij}(p,p')=\frac{2}{\pi}\int_0^\infty r^2dr\; j_0(pr)V_{ij}(r)j_0(p'r) \, ,
\label{eq20} 
\end{equation}
and the two-body potentials consist of an attractive and a repulsive
Yukawa term as in Eq.~\eqref{eq21}. The parameters of the $\Xi N$ channels 
were obtained by fitting the low-energy data as given 
in the most recent update of the strangeness $-2$ Nijmegen ESC08c 
potential~\cite{Nag15}. Besides, as mentioned above, the HAL QCD Collaboration~\cite{Sas18} has recently derived 
a potential for the $(i,j)=(0,0)$ $\Lambda\Lambda -\Xi N$ channel with almost
physical quark masses. Thus, we have performed the calculation with both models
for the $(i,j)=(0,0)$ $\Lambda\Lambda -\Xi N$ channel, Nijmegen ESC08c~\cite{Nag15} 
and HAL QCD~\cite{Sas18}.
The low-energy data and the parameters of the different $\Xi N$ interactions
are given in Table~\ref{tabx3}.

With $N=15$ generalized Gaussians in Eq.~\eqref{Rkr} we have obtained a $\Xi^- p nn$ bound state of $14{.}43$ MeV with
the $(i,j)=(0,0)$ HAL QCD interactions and $10.78$ MeV with the Nijmegen potentials\footnote{Note that 
the mass of $^4{\rm He}$ changes by 0.24 MeV from $N=15$ to $N=25$, so the result
is fully converged.}. In both cases, the $(I,J)=(1,0)$ $\Xi^- p nn$ state lies below the 
lowest two-body threshold, $\Xi^- t$. Such state would decay to the $\Lambda\Lambda nn$ channel 
with a very small width as shown in Sec.~\ref{secIIc} and Ref.~\cite{Gar18}.
The results are in close agreement with those obtained with the separable potential
three-body model shown in Table~\ref{tab3}. In all models the binding is larger than that of the
tritium and a slightly deeper bound state is obtained when using the HAL QCD interactions for the two-body
coupled channel $(i,j)=(0,0)$. By including the 
Coulomb $\Xi^- p$ potential the binding energies are increased roughly by $0{.}75$ MeV with the 
HAL QCD interaction and $0{.}53$ MeV with the Nijmegen potentials, driving to final binding
energies of $15{.}18$ MeV and $11{.}31$ MeV, respectively. 

\section{Outlook}

It has been suggested in Ref.~\cite{Ble19} that some of the structures observed 
in the correlated $\pi^- - \pi^-$ momenta by the BNL AGS-E906
experiment~\cite{BNLAGS}, aiming to produce and study double hypernuclei through
a $(K^-,K^+)$ reaction on $^9$Be,
could result from the decays of a $^4_{\Lambda\Lambda}$n double hypernucleus.
We have studied the coupled $\Lambda\Lambda nn-\Xi^- pnn$ system 
to check if the inclusion of channel coupling is able
to bind the $\Lambda\Lambda nn$ system.
We have used two different approaches. The first one is a separable potential 
three-body model of the coupled $\Lambda\Lambda nn - \Xi^- pnn$ system tuned
to the known experimental data that allows us to evaluate the $\Xi^- t$ binding
energy and its decay width to $\Lambda\Lambda nn$. The second one is a 
generalized Gaussian variational method based on realistic two-body
interactions tuned in the known two-, three- and four-body systems experimental 
data. 

With the available two-body interactions that are adjusted to describe what is known 
about the two- and three-baryon subsystems, neither a $\Lambda\Lambda nn$ 
bound state nor a resonance is obtained. However, we have
found a $\Xi^- t$ quasibound state with quantum numbers $(I,J)=(1,0)$ 
above the $\Lambda\Lambda nn$ threshold.
The stability of the state is increased by considering the Coulomb potential.
The different approaches to the $\Lambda\Lambda - \Xi N$ interaction
drive to similar results, the weakness of the 
$\Lambda\Lambda - \Xi N$ transition potential explaining the narrow width of the
$\Xi^- t$ quasibound state. Finally, we have calculated the $\Xi^-t$ scattering 
length, which may be useful in the calculation of the energy shift of the 
atomic levels of the $\Xi^-t$ atom.

\section{Acknowledgments} 
This work has been partially funded by COFAA-IPN (M\'exico), 
by Ministerio de Econom\'\i a, Industria y Competitividad 
and EU FEDER under Contract No. FPA2016-77177.

\end{document}